\documentclass[aps,twocolumn,nofootinbib,groupedaddress]{revtex4-1}

\usepackage[english]{babel}
\usepackage{graphicx}
\usepackage{amsmath,amssymb,amsbsy,amstext, amsthm, simplewick, amsfonts}
\usepackage{dcolumn}

\usepackage{hyperref}
\hypersetup{
    colorlinks=true,
    linkcolor=purple,
    filecolor=purple,      
    urlcolor=purple,
    citecolor=purple
    }
    
\usepackage{comment}
\usepackage{xcolor}
\usepackage{braket}
\usepackage{tensor}
\usepackage{ulem}
\usepackage{dsfont}
\usepackage{bm}

\usepackage{tikz}
\usetikzlibrary{decorations.markings}
\usetikzlibrary{shapes.misc}
\usetikzlibrary{arrows}
\usetikzlibrary{fit, chains}

\def \d {\mathrm{d}}

\def \x {\bm{x}}

\def \k {\bm{k}}

\def \q {\bm{q}}

\def \a {\textrm{a}}

\def \mut {\tilde{\mu}}

\def \C {\mathcal{C}}
\def \D {\mathcal{D}}
\def \E {\mathcal{E}}

\def \L {\mathcal{L}}
\def \G {\mathcal{G}}
\def \I {\mathcal{I}} 

\def \N {\mathcal{N}}
\def \O {\mathcal{O}}
\def \P {\mathcal{P}}

\def \H {\mathcal{H}}
\def \V {\mathcal{V}}

\def \aa {{\sf{a}}}

\def \SO {\mathrm{SO}}
\def \dS {\mathrm{dS}}

\def \EAdS {\mathrm{EAdS}}

\def \KLF {\mathrm{KLF}}

\usepackage{colortbl}
\definecolor{pyblue}{RGB}{31, 119, 180}
\definecolor{pyred}{RGB}{214, 39, 40}
\definecolor{pygreen}{RGB}{44, 160, 44}

\begin{document}


\title{De Sitter Momentum Space}

\author{Nathan Belrhali$^{\rm K}$}
\author{Arthur Poisson$^{\rm K}$}
\email{poisson@iap.fr}
\author{S\'ebastien Renaux-Petel$^{\rm K}$}
\author{Denis Werth$^{\rm K, L, F}$}

\affiliation{$^{\rm K}$Institut d'Astrophysique de Paris, CNRS, Sorbonne Universit\'e, FR-75014, France \\
$^{\rm L}$Max-Planck-Institut f\"ur Physik, Werner-Heisenberg-Institut, Garching bei M\"unchen, D-85748, Germany \\
$^{\rm F}$Max Planck-IAS-NTU Center for Particle Physics, Cosmology and Geometry}

\begin{abstract}
We construct a natural and nonperturbative momentum space for quantum field theory on $(d+1)$-dimensional de Sitter (dS) spacetime in the Poincar\'e slicing, adapted to early Universe cosmology. In particular, we identify the dS frequency as the unitary-representation label of the $\dS$ isometry group $\SO(1, d+1)$. By diagonalizing the quadratic Casimir together with spatial translations, we provide a harmonic expansion of operators in what we call the Kontorovitch-Lebedev-Fourier (KLF) space.
This momentum space shares many structural properties with its Minkowski counterpart, for instance: equations of motion reduce to algebraic equations, and the quadratic dynamics provides a simple propagator analogous to flat space. We reformulate the perturbative computation of in-in correlators in KLF momentum space, showing from first principles how time integrals turn into frequency-space integrals over meromorphic functions.
We show how our construction  streamlines computations, naturally accommodates the contributions from principal and complementary series in the K\"all\'en-Lehmann spectral decomposition of composite operators, and leads to a group-theoretical method to evaluate loop momentum integrals.
\end{abstract}

\maketitle


{\bf Introduction.} 
Among the three maximally symmetric solutions of the vacuum Einstein equations, de Sitter (dS) spacetime occupies a distinguished role: it describes both the early inflationary Universe and its late-time accelerated expansion. Yet despite this privileged status, quantum field theory (QFT) in dS remains strikingly incomplete and poorly understood. 

Unlike anti-de Sitter space, dS admits no holographic dual, obstructing a nonperturbative definition of its quantum dynamics, let alone rendering its strongly coupled regime inaccessible~\cite{Strominger:2001pn,Witten:2001kn}. The absence of a global timelike Killing vector prohibits a conserved notion of energy, and continuous particle production obstructs a conventional $S$-matrix interpretation, reflecting the absence of asymptotic future states (see~\cite{Marolf:2012kh,Melville:2023kgd,Donath:2024utn,Melville:2024ove} for proposals). These conceptual limitations are mirrored by severe technical obstacles for perturbation theory. Late-time correlators are difficult to compute and their analytic structure is not yet fully understood, especially at the loop level (see e.g.~\cite{DiPietro:2021sjt,Xianyu:2022jwk,Chakraborty:2023qbp,Qin:2023bjk,Qin:2023nhv} for recent progress). In the conformal-time and spatial-momentum representation, the conventional computation of cosmological correlators parallels that of flat-space scattering amplitudes in {\it configuration space}, involving integrals over products of Bessel-type special functions~\cite{Chen:2017ryl}. Just as flat-space momentum space reorganizes these expressions into simple rational propagators, it is natural to seek an analogous momentum-space formulation for dS correlators, dual to position space.

These difficulties share a common origin: the usual Fourier transform in time and space does not diagonalize the dS isometries. As a result, symmetry constraints, dynamical evolution and analytic structure of observables remain entangled.

In this letter, we show that many of these obstacles are artifacts of representation. We construct a nonperturbative momentum-space formulation of QFT in dS, adapted to its symmetries. We call this space, dual to position-space Poincar\'e coordinates $(\tau,\x)$ and labeled by $(\mu,\k)$, the {\it Kontorovich-Lebedev-Fourier} (KLF) space, where the unitary-representation label $\mu$ provides the appropriate quantum number for the dS frequency.

Of course, a dS momentum space for global coordinates in Euclidean signature, after complexifying (global) time, dates back to many earlier works~\cite{Higuchi:1986wu,Bros:1990cu,Bros:1995js,Marolf:2010zp,Marolf:2010nz,Higuchi:2010xt}. This formulation provides a nonperturbative definition of QFT free of infrared divergences, with a uniquely defined (Euclidean) Bunch-Davies vacuum. 
There, the theory is defined on the $(d+1)$-dimensional sphere and decomposes into spherical harmonics. However, its compactness obscures the dynamics, which can only be accessed via subtle analytic continuation.

By contrast, in the KLF momentum space, dS symmetries, dynamics, and analyticity become simultaneously transparent, equations of motion reduce to algebraic relations, and the Feynman propagator takes the same form as in flat space. We show that perturbative calculations of newly defined (amputated) Schwinger-Keldysch (SK) correlators are free of nested time integrals and instead take the form of spectral integrals over frequencies. The resulting integrands being universally meromorphic, this enables correlators to be evaluated by simply collecting residues. At loop level, we show how our group-theoretical construction streamlines and gives a transparent physical meaning to the momentum integration in a simple example.
We provide extensive details and applications in a companion paper~\cite{Belrhali:2026rkn}.


\vskip 4pt
{\bf dS spacetime.}
For the sake of generality and to allow the use of dimensional regularization, throughout this letter, we consider $(d+1)$-dimensional dS spacetime $\dS_{d+1}$, defined as an hyperboloid embedded in $(d+2)$-dimensional Minkowski spacetime $\mathbb{M}^{1, d+1}$:
\begin{equation}
\label{eq: dS def}
     \eta_{AB} X^A X^B= H^{-2} \,,
\end{equation}
where $A, B = 0, 1, \ldots, d+1$ and $H^{-1}$ is the dS curvature radius. For cosmological purposes, it is enough to restrict our attention to the half of $\dS$ space covered by  Poincar\'e coordinates, such that $X^0 = (2 H \tau)^{-1}(\tau^2-\x^2-1)\,, X^i = -(H\tau)^{-1} x^i\,, X^{d+1} = (2 H \tau)^{-1}(\x^2-\tau^2-1)$, where $-\infty<\tau< 0$ is conformal time and $x^i$, $i=1, \ldots, d$ are spatial coordinates. With these, the metric induced on the hyperboloid~\eqref{eq: dS def} takes the cosmological form $ \d s^2 =(H\tau)^{-2}\left(-\d\tau^2 + \d\x^2\right)$, which describes an accelerating spacetime with constant Hubble rate $H$ and scale factor $-(H\tau)^{-1}$. As can be read off from the definition~\eqref{eq: dS def}, $\dS_{d+1}$ is left invariant by the $(d+2)$-dimensional Lorentz group $\SO(1,d+1)$. Instead, constant-$\tau$ spatial hypersurfaces are $d$-dimensional Euclidean planes $\mathbb{R}^d$, which are only invariant under the action of the Euclidean group $\mathbb{E}_d \subset \SO(1,d+1)$, composed of translations and rotations in $d$ dimensions.


\vskip 4pt
{\bf Hilbert space of QFT in dS.} 
The generators of the symmetry group $\SO(1,d+1)$ are realized as Hermitian operators acting on the Hilbert space $\H$. Therefore, the latter decomposes into Unitary Irreducible Representations (UIRs) of $\SO(1, d+1)$: $\H = \bigoplus_\mu \V_{\mu}$ where $\mu$ is an abstract label of the UIRs and $\V_{\mu}$ are the corresponding invariant subspaces~\cite{Bargmann:1948ck}. The UIRs are classified according to the eigenvalues $M^2_{\mu}$ of the quadratic Casimir operator $\C = -\frac{1}{2}J^{AB} J_{AB}$, where $J_{AB}=-J_{BA}$ are the generators of the $(d+2)$-dimensional Lorentz algebra. The Casimir acts as a multiple of the identity when restricted to each $\V_{\mu}$,  $\C = \bigoplus_\mu M^2_{\mu} \mathds{1}^{\mu}$, where the eigenvalues $\Delta(d-\Delta) + S(S+d-2)$ generally depend on a complex conformal dimension $\Delta$ and a representation $S$ of the rotation subgroup $\SO(d)$ (see~\cite{Sun:2021thf} for a review). For simplicity, we only consider scalar representations here and set $S=0$. With the parametrization $\Delta =\tfrac{d}{2}+i \mu$, the Casimir eigenvalues of interest read $M_{\mu}^2 = \mu^2 + \tfrac{d^2}{4}$, whose form makes manifest the $\mathbb{Z}_2$ shadow symmetry $\mu \leftrightarrow -\mu$. As the generators are Hermitian, the Casimir eigenvalues are real, implying that the parameter $\mu$ is either real or purely imaginary. Discarding the ever-present trivial representation, the possible scalar UIRs of $\SO(1, d+1)$ fall into three categories: (i) the principal series $\P_{\mu}$ with $\mu \in \mathbb{R}$, (ii) the complementary series $\C_{\mu}$ with $i\mu \in [-\tfrac{d}{2}, \tfrac{d}{2}]$, and (iii) the type-I exceptional series $\E_{\mu}$ with $i\mu = \tfrac{d}{2} + k$ and $k\in \mathbb{Z}_{\geq 0}$. The free bulk theory whose single-particle Hilbert space realizes these representations corresponds, respectively, to heavy fields $m > \tfrac{d}{2}H$, light fields $m\leq \tfrac{d}{2}H$, and scalar fields with extended shift symmetries, see e.g.~\cite{Bonifacio:2018zex} for the latter aspect.

To determine a basis of the Hilbert space, we follow a standard procedure: we identify the largest set of commuting generators and diagonalize them in a common orthogonal basis together with the Casimir operator. To make contact with flat space and cosmological observables, we do so in a way that takes advantage of the $\mathbb{E}_d$ symmetry of spacelike slices. To this end, it is convenient to define the following change of basis in the Lie algebra:
\begin{equation}
\label{eq: de Sitter generator conformal split}
\begin{aligned}
    D = J_{0, d+1} &\,, \quad M_{ij} = J_{ij} \,, \\ 
    P_i = J_{d+1, i}+J_{0, i} &\,, \quad K_i = J_{d+1, i} - J_{0, i} \,.
\end{aligned}
\end{equation}
The generators $D, P_i, K_i$ and $M_{ij}$ correspond to dilatation, spatial translations, special conformal transformations (sometimes referred to as dS boosts), and spatial rotations, respectively. 
Diagonalizing the translation operators yields the following momentum basis for the Hilbert space:
\begin{equation}
\label{eq:momentum states}
\begin{aligned}
    \C\ket{\mu, \k} = M_{\mu}^2 \ket{\mu, \k} \,,\quad P_i \ket{\mu, \k} = k_i \ket{\mu, \k} \,,
\end{aligned}
\end{equation}
which are identified with single-particle states that can be constructed from the vacuum. 


\vskip 4pt
{\bf Kontorovich-Lebedev-Fourier space.} 
The central quantities of interest are correlation functions evaluated in a given state. In inflationary cosmology, one is interested in the Bunch-Davies vacuum state $\ket{\Omega}$, defined by the absence of particles in the asymptotic past $\tau\to-\infty$.  In the SK (in-in) formalism, this is implemented by performing a path-integral on a first branch from $-\infty(1- i \epsilon)$ to late times, then on a second branch backwards to $-\infty(1+ i \epsilon)$ in the complex time plane~\cite{Weinberg:2005vy}.
This path-integral is performed over field configurations that are square-integrable. It is therefore essential to characterize this functional space. For this, let us introduce the following wavefunction:
\begin{equation}
\label{eq:wavefunctions}
    \Phi^{\mu}_{\k,\O}(\tau,\x) \equiv \bra{\Omega}\O(\tau,\x)\ket{\mu,\k} \,,
\end{equation}
where $\O(\tau,\x)$ stands for any local Hermitian operator. Using that the action of the generators~\eqref{eq: de Sitter generator conformal split} on $\O$ is given by the differential operator associated with the corresponding Killing vectors, one deduces that the Casimir acts on $\O$ as the Laplace-Beltrami operator on $\dS_{d+1}$:
\begin{equation}
\begin{aligned}
   &[\C, \O(\tau, \x)] = H^{-2}\Box_{\dS}\O(\tau,\x) \\
    &=-\left(\tau^2\partial_\tau^2-(d-1)\tau\partial_\tau-\tau^2\partial_i^2\right)\O(\tau,\x) \,.
\end{aligned}
\end{equation}
This, together with the definition of the momentum basis~\eqref{eq:momentum states}, implies that the wavefunction~\eqref{eq:wavefunctions} is an eigenfunction of the Killing vectors associated with $\C$ and translations $P_i$. However, these operators are not self-adjoint on the tilted axis of the SK path-integral contour. Instead, they are self-adjoint once the two SK branches are Wick-rotated to the imaginary axis:
\begin{equation}
    \int \displaylimits_{-\infty(1\mp i \epsilon)}^0\d\tau f(\tau) = \mp i\int \displaylimits_0^\infty\d z f(e^{\pm i \frac{\pi}{2}}z) \,,
\end{equation}
which maps $\dS_{d+1}$ geometry to the one of Euclidean anti-de Sitter $\EAdS_{d+1}$
(see, e.g.~\cite{Anninos:2014lwa,Sleight:2021plv,DiPietro:2021sjt,Loparco:2023rug,Chowdhury:2023arc} for relevant works in that context). It follows that the analytic continuation of the wavefunction satisfies the differential equations:
\begin{subequations}
\label{eq: diff eqs}
    \begin{align}
    \label{eq: diff eq Casimir}
        H^{-2}\Box_{\EAdS} \Phi_{\k,\O}^{\mu}(z, \x) &= -M_{\mu}^2 \Phi_{\k,\O}^{\mu}(z, \x) \,, \\
    \label{eq: diff eq Translation}
        \partial_i \Phi_{\k,\O}^{(\mu)}(z, \x) &= -ik_i \Phi_{\k,\O}^{\mu}(z, \x) \,,
    \end{align}
\end{subequations}
with $H^{-2}\Box_{\EAdS} \equiv \left(z^2 \partial_z^2 - (d-1)z\partial_z + z^2 \partial_i^2\right)$. The set of equations~\eqref{eq: diff eq Casimir}-\eqref{eq: diff eq Translation} define a well-posed Sturm-Liouville problem on $L^2\left[\EAdS_{d+1},\d z\d^d\x\sqrt{-g}\right]$, whose solution reads $ \Phi^{\mu}_{\k,\O}(z,\x)=c_\O(\mu)  \Phi^{(\mu)}_{\k}(z,\x)$, where $c_\O(\mu)$ is a theory-dependent normalization, and
\begin{equation}
\label{eq: expression Harmonic Function}
    \Phi^{(\mu)}_{\k}(z,\x) = \frac{H^{\frac{d+1}{2}}}{\sqrt{\pi}}e^{-i\k\cdot\x}z^{\frac{d}{2}}K_{i\mu}(k z) \,,
\end{equation}
where $K_{i\mu}(k z)$ is the modified Bessel function of second type, which is real. Since $\Box_{\EAdS}$ and $\partial_i$ are self-adjoint, the functions~\eqref{eq: expression Harmonic Function} form an orthonormal set with respect to the $L^2$ inner product.

Importantly, we now show that the harmonic functions~\eqref{eq: expression Harmonic Function} constitute a complete basis of $L^2\left[\EAdS_{d+1},\d z\d^d\x\sqrt{-g}\right]$ for $\mu \in \mathbb{R}$. For this, let us introduce the shifted Euclidean Casimir operator in spatial Fourier space
\begin{equation}  \Delta_z=z^{d+1}\frac{\d}{\d z}\left(\frac{1}{z^{d-1}}\frac{\d}{\d z}\right)-(kz)^2+\frac{d^2}{4}\,,
\end{equation}
which is self-adjoint on $L^2[\mathbb{R}_+,\d z/z^{d+
1}]$. For $\lambda$ complex, we now define the Green function $G^{\lambda}(z,z')$, such that $(\Delta_z-\lambda)G^{\lambda}(z,z')=z'^{d+1}\,\delta(z-z')$, and with vanishing boundary conditions appropriate for $L^2$. It can be built from the homogeneous solutions $z^{\frac{d}{2}}K_{\sqrt{\lambda}}(k z)$ and $z^{\frac{d}{2}}I_{\pm\sqrt{\lambda}}(k z)$, where $I$ is the modified Bessel function of first type, and reads
\begin{equation}
\label{eq: Green Function 1}
    G^{\lambda}(z,z') = -(z z')^{\frac{d}{2}}\left\{\begin{array}{cc}
        I_{\sqrt{\lambda}}(k z)K_{\sqrt{\lambda}}(k z') & \text{for}\; z<z' \\
        I_{\sqrt{\lambda}}(k z')K_{\sqrt{\lambda}}(k z) & \text{for}\; z'<z 
    \end{array}\right. \,.
\end{equation}
Since this is an analytic function of the parameter $\lambda$ in the cut plane $\mathbb{C}\setminus\mathbb{R}_-$, it can be rewritten as a dispersive integral over its discontinuity, yielding:
\begin{equation}\label{eq:disp int}
    G^{\lambda}(z,z') = \int\displaylimits^0_{-\infty}\frac{\d\lambda'}{2i\pi}\frac{\text{Disc}_{\lambda'}(G^{\lambda'}(z,z'))}{\lambda'-\lambda} \,.
\end{equation}
The discontinuity can be deduced from Eq.~\eqref{eq: Green Function 1} using the connection formula among Bessel functions~\cite{NIST:DLMF}. After changing variable $\lambda=-\mu^2$, one finds
    $G^\lambda(z,z')=-\frac{(zz')^{\frac{d}{2}}}{\pi}\int \displaylimits^\infty_{-\infty}\d\mu\, \N_\mu\frac{K_{i\mu}(kz)K_{i\mu}(kz')}{\mu^2-\mu^2_\lambda}$, where we define $\N_\mu \equiv \frac{\mu}{\pi}\sinh(\pi\mu)$ which identifies to the de Sitter density of states. Acting with $\Delta_z-\lambda$ on this expression, one deduces
\begin{equation}
    \frac{(z z')^{\frac{d}{2}}}{\pi}\int\displaylimits_{-\infty}^\infty\d\mu\,\N_\mu\, K_{i\mu}(k z)K_{i\mu}(k z')= z^{d+1}\delta(z-z')\,.
\end{equation}
This shows that the functions $z^{\frac{d}{2}}K_{i\mu}(k z)$, for $\mu$ real, form a complete basis of $L^2\left[\mathbb{R}_+,\d z/z^{d+1}\right]$, whose decomposition is known to mathematicians as the Kontorovich-Lebedev transform~\cite{Kontorovich1938, Lebedev1946}.
 
The spatial component  of the harmonic functions~\eqref{eq: expression Harmonic Function} being the usual $d$-dimensional Fourier plane waves, this motivates introducing a $(d+1)$-dimensional momentum space that we call the {\it Kontorovich-Lebedev-Fourier} (KLF) space. Any square-integrable function on $\EAdS_{d+1}$ admits the following decomposition:
\begin{equation}
\label{eq: KLF transform}
\begin{aligned}
    f(z,\x) &= \int_{\KLF} \;\Phi^{\mu}_{\k}(z,\x)\;f^{\mu}_{\k}\;,\\
    f^{\mu}_{\k} &= \int_{\EAdS}\left[\Phi^{\mu}_{\k}(z,\x)\right]^*\, f(z,\x) \,,
\end{aligned}
\end{equation}
where the integration measures are
\begin{equation}
    \begin{aligned}
        \int_{\KLF} &\equiv \int \displaylimits_{-\infty}^{+\infty}\d\mu \, \N_{\mu}\int_{\mathbb{R}^d}\frac{\d^d\k}{(2\pi)^d} \,, \\
        \int_{\EAdS} &\equiv \int \displaylimits_0^\infty \frac{\d z}{(Hz)^{d+1}}\int_{\mathbb{R}^d}\d^d\x \,.
    \end{aligned}
\end{equation}
The reader may wonder why the KLF harmonic functions~\eqref{eq: expression Harmonic Function} are not themselves square-integrable. This is in fact not different from plane waves not being square-integrable either and still be the correct basis for the Fourier transform. This situation arises when the operator to be diagonalized is unbounded, in which case the proper mathematical structure is not the Hilbert space but the Gelfand triple~\cite{Gelfand-Vilenkin,Maurin}. 
\vskip 4pt
The objects we encounter in physical situations need not be square-integrable, and the KLF transform~\eqref{eq: KLF transform} should thus be extended to more general functions.
To this aim, we introduce the generalized orthogonality relation:
\begin{equation}
\label{eq:generalized-delta}
    \begin{aligned}
        \int_{\EAdS}\left[\Phi^{\mu}_{\k}(z,\x)\right]^*\Phi^{\alpha}_{\k'}(z,\x) \\
        =\frac{(2\pi)^d}{\N_{\mu}}\delta^{(d)}(\k-\k')\hat{\delta}_{\alpha}(\mu) \,,
    \end{aligned}
\end{equation}
where the generalised delta function $\hat{\delta}_{\alpha}(\mu)$ is defined for any $\alpha\in \mathbb{C}$ as
$\int_{\mathbb{R}}\d\mu\;\hat{\delta}_{\alpha}(\mu)f^{(\mu)}_{\k} =\frac12(f^{(\alpha)}_{\k}+f^{(-\alpha)}_{\k})$, and $f^{(\alpha)}_{\k}$ is defined as the analytical continuation of the KLF function $f^{(\mu)}_{\k}$.
For $\alpha$ real, this reduces to $\hat{\delta}_{\alpha}(\mu) =|\alpha|\delta(\mu^2-\alpha^2)$. This is a key aspect of our formalism: all computations can be carried out first for fields in the principal series, free of any difficulty like infrared divergences; more general physical results then simply follow by analytical continuation.
In what follows, for notational simplicity, we will write the KLF delta function as $\delta(^{\mu\,\mu'}_{\k\,\k'})\equiv(2\pi)^d\delta^{(d)}(\k-\k') \;\N_{\mu}^{-1}\,\hat{\delta}_{\mu'}(\mu)$. 


\vskip 4pt
{\bf KLF-space correlators.} 
KLF-space correlators are defined for real $\mu_i$ as:
\begin{equation}
\label{eq: definition G via Z}
    \mathcal{G}_{\aa_n \ldots \aa_n} \left(^{\mu_1...\mu_n}_{\k_1...\k_n}\right) = \prod_{i=1}^n\frac{(2\pi)^d}{\N_{\mu_i}}\frac{\delta Z[J_+,J_-]}{\delta J_{\aa_i,\k_i}^{\mu_i}}  \Bigg\rvert_{J_\pm=0} \,,
\end{equation}
where the functional generator reads 
\begin{equation}
\label{eq: functional Generator KLF}
    Z[J_+,J_-] \equiv \int\D\varphi_\pm e^{i S_+ - i S_- + \int_{\KLF}(\varphi_+J_+ +\varphi_-J_-)} \,,
\end{equation}
and the integral runs over $L^2\left[\EAdS_{d+1},\d z\d^d\x\sqrt{-g}\right]$ functions. As explained above, the action integrals are taken over the two SK branches:
\begin{equation}
\label{eq:action generic}
    S_\pm[\varphi] = \hspace*{-0.4cm}\int \displaylimits^0_{-\infty(1\mp i \epsilon)}\hspace*{-0.4cm}\d\tau\d^d\x\sqrt{-g}\,\L(\varphi,\partial_\mu\varphi)\;,
\end{equation}
with a sewing boundary condition at late time $\varphi_+(\tau_0)=\varphi_-(\tau_0)$. The functional generator can be explicitly evaluated for a Gaussian action
\begin{equation}
\label{eq: free action scalar}
    S_0 = \frac{1}{2}\int \frac{\d\tau\,\d^d\x}{(-H\tau)^{d-1}}\left[\left(\partial_\tau\varphi\right)^2-\left(\partial_i\varphi\right)^2 -\frac{m_{\varphi}^2}{H^2 \tau^2}\varphi^2\right] \,,
\end{equation}
which, after analytically continuing to $\EAdS$ and decomposing onto the KLF harmonic basis, yields
\begin{equation}
    \pm i S_{0,\pm}[\varphi] = -\frac{H^2 e^{\pm\frac{i(d-1)\pi}{2}}}{2}\int_{\KLF}\varphi^{\mu}_{\k}(\mu^2-\mu_{\varphi}^2)\varphi^{\mu}_{-\k} \,.
\end{equation}
Following~\cite{Weinberg:2005vy}, one can complete the square in the path-integral~\eqref{eq: functional Generator KLF} to find the four KLF-space propagators. For principal series fields, it reads:
\begin{equation}
\label{eq: propagator matrix}
    \bm{\mathcal{G}}_{\mu_\varphi}(\mu) = \left(\begin{array}{cc}
    \frac{e^{-\frac{i\pi(d-1)}{2}}}{(\mu^2-\mu_{\varphi}^2)_{i\epsilon}} & \frac{\hat{\delta}_{\mu_{\varphi}}(\mu)}{\N_{\mu}} \\
    \frac{\hat{\delta}_{\mu_{\varphi}}(\mu)}{\N_{\mu}} & \frac{e^{+\frac{i\pi(d-1)}{2}}}{(\mu^2-\mu_{\varphi}^2)_{-i\epsilon}}
    \end{array}\right) \,,
\end{equation}
where we defined the reduced propagator matrix as $\boldsymbol{\mathcal{G}}_{\mu_\varphi}\left(^{\mu\,\mu'}_{\k\,\k'}\right) \equiv H^{-2} \delta(^{\mu\,\mu'}_{\k\,-\k'})\boldsymbol{\mathcal{G}}_{\mu_\varphi}(\mu)$. For non-principal $\varphi$, the diagonal components of~\eqref{eq: propagator matrix} should include an additional contribution from the $\V_{\mu_\varphi}$ UIR \cite{unpublishedkey}. In~\eqref{eq: propagator matrix}, the suitable $i\epsilon$ prescription that encodes particle propagation reads~\cite{Melville:2024ove, Werth:2024mjg}
\begin{equation}
     \frac{1}{(\mu^2-\mu_{\varphi}^2)_{i\epsilon}} \equiv \frac{1}{2\sinh(\pi\mu_{\varphi})}\left[\frac{e^{+\pi\mu_{\varphi}}}{\mu^2-\mu_{\varphi}^2+i\epsilon}-\frac{e^{-\pi\mu_{\varphi}}}{\mu^2-\mu_\varphi^2-i\epsilon}\right] \,.
\end{equation}
The diagonal terms in~\eqref{eq: propagator matrix} are the equivalent of the Minkowski Feynman propagators in momentum space, $\G(p) =\pm i/(-p^2-m_{\varphi}^2 \pm i\epsilon)$. 
\vskip 4pt
Following the usual path-integral derivation, it is straightforward to write Feynman rules for KLF correlators. Each vertex comes with a SK $\pm$ index, propagators are given by~\eqref{eq: propagator matrix}, and polynomial interactions $\L = -\lambda \phi^n/n!$ contribute as

\begin{equation}
\begin{aligned}
    &\begin{tikzpicture}[baseline={(0,0)}]
    \draw[thick, black] (0, 0) to (-1,1) node[above] {\scriptsize $\mu_1,\k_1$};
    \draw[thick, black] (0, 0) to (1,1) node[above] {\scriptsize $\mu_n,\k_n$};
    \draw[thick, black] (0,0) to (-0.33,1) node[above, xshift=1mm] {\scriptsize $\mu_2,\k_2$};
    \node[black] at (0.2,0.7) {$\ldots$};
    \begin{scope}
        \clip (0,0) circle(2pt);
        \fill[black] (0,-2pt) rectangle (-2pt,2pt);
        \fill[white] (0,-2pt) rectangle (2pt,2pt);
    \end{scope}
    \draw[color=black] (0,0) circle (2pt) node[below] {\scriptsize$\aa=\pm$};
    \end{tikzpicture}
    &\hspace*{-0.8cm}= - i\aa\lambda\frac{H^{\frac{(n-2)(d+1)}{2}}e^{\aa\frac{i \pi d}{2}}}{\pi^\frac{n}{2}}\mathcal{I}^{\mu_1 \ldots \mu_n}_{k_1 \ldots k_n} \,,\\
    \end{aligned}
    \label{KLF-vertex}
\end{equation}
where we kept the spatial momenta-conserving delta function implicit and the $n$-point vertex functions read
\begin{equation}
    \mathcal{I}^{\mu_1 \ldots \mu_n}_{k_1 \ldots k_n} = \int \displaylimits^\infty_0\d z\; z^{\frac{d(n-2)}{2}-1}\prod_{j=1}^n K_{i\mu_j}(k_j z) \,.
    \label{eq: def vertex function}
\end{equation}
Instead of the energy-conserving delta function one would get in flat space, the breaking of time translation invariance in $\dS$ entails a complicated time-dependence of the harmonic function~\eqref{eq: expression Harmonic Function}, and hence the nonconservation of $\mu$'s at vertices. The vertex functions $\mathcal{I}$ are universally meromorphic, see e.g.~\cite{Bzowski:2013sza,Bzowski:2015yxv} for $n=3$, and constitute the fundamental building blocks of KLF correlators. Using these rules, one can write the expressions of amputated KLF correlators $\G^A_{\{\aa_E^V\}}(^{\mu_1 \ldots \mu_n}_{\k_1 \dots \k_n})$, where propagators associated with external lines are removed and $\{\aa_E^V\}$ are the $n_E$ external vertices.


\vskip 4pt
{\bf From KLF to cosmological correlators.} 
Amputated KLF correlators are simply related to $n$-point cosmological correlators by means of the reduction formula: 
\begin{equation}
\label{eq: KLF correlators in real time}
    \braket{\prod_{i=1}^n\varphi_{\aa_i,\k_i}^i(\tau_0)} = \E_{\{\aa_E^V\}}(\tau_0)\G^A_{\{\aa^V_E\}}\left(^{\mu_{\varphi_1}\ldots\mu_{\varphi_n}}_{\k_1\;\ldots\;\k_n}\right)\,,
\end{equation}
where we omitted the momentum-conserving delta function to avoid clutter.
Here, each field $\varphi^i$ has frequency $\mu_{\varphi_i}$, and each $\aa_j\in\{\aa_E^V\}$ is attached to $n_j$ external legs whose contributions are $\E_{\{\aa_E^V\}}(\tau_0) \equiv \prod_{j=1}^{n_E}\prod_{\ell=1}^{n_j}H^{-2}\Phi^{\mu_{\varphi_\ell}}_{\k_\ell}\left(-\tau_0 e^{-\frac{i\aa_j\pi}{2}},\boldsymbol{0}\right)$.
Physical correlators are then obtained by summing diagrams over all the possible $\{\aa_E^V\}$ configurations. Let us highlight that, although KLF correlators~\eqref{eq: definition G via Z} are initially defined for real $\mu_i$, they can be straightforwardly analytically continued to arbitrary complex $\mu_i$. As a simple example, the $s$-channel ($++$) amputated KLF diagram arising from $-\frac12 g \varphi^2 \chi$ reads
\begin{widetext}
\begin{equation}
    \begin{aligned}
        \G^A_{++}\left(^{\mu_1 \mu_2 \mu_3 \mu_4}_{\k_1 \k_2 \k_3 \k_4}\right) \equiv 
        &\begin{tikzpicture}[baseline={(0,0)}]
            \draw[thick,black] (-1.,0.6) to (-0.5,0);
            \draw[thick,black] (-1.,-0.6) to (-0.5,0);
            \draw[thick,black] (1.,0.6) to (0.5,0);
            \draw[thick,black] (1.,-0.6) to (0.5,0);
            \draw[thick,pyblue] (-0.5,0) to (0.5,0) node[midway, below=0.5mm] {\scriptsize $\mu,\k_I$};
            \filldraw[color=black,fill=black] (-0.5,0) circle (2pt);
            \filldraw[color=black,fill=black] (0.5,0) circle (2pt);
            \node[black] at  (-1.14,1) {\scriptsize $\mu_1,\k_1$};
            \node[black] at  (-1.14,-1) {\scriptsize $\mu_2,\k_2$};
            \node[black] at  (1.14,1) {\scriptsize $\mu_3,\k_3$};
            \node[black] at  (1.14,-1) {\scriptsize $\mu_4,\k_4$};
        \end{tikzpicture}
        & = (-i g)^2 \frac{H^{d-1}e^{i\pi d}}{\pi^3} \int \displaylimits^{+\infty}_{-\infty}\d\mu\mathcal{N}_{\mu}\frac{\mathcal{I}^{\mu_1\mu_2\mu}_{k_1k_2k_I}\mathcal{I}^{\mu\;\mu_3\mu_4}_{k_Ik_3k_4}}{(\mu^2-\mu_{\chi}^2)_{i\epsilon}} \,,
    \end{aligned}
    \label{eq:exchange}
\end{equation}    
\end{widetext}
The exchanged momentum $\k_I$ is fixed by $d$-momentum conservation, but, the nonconservation of the dS frequency implies a spectral integral. This provides a first-principle symmetry-based derivation of the spectral decomposition of correlators~\cite{Melville:2024ove, Werth:2024mjg}. Specifying to external conformally coupled fields $\mu_j = i/2$ for $j=1, \ldots, 4$, the frequency-space integral can be finished by closing the contour at infinity and picking up residues: 
\begin{widetext}
\begin{equation}
    \begin{aligned}
       \G^A_{++}\left(^{\mu_1 \mu_2 \mu_3 \mu_4}_{\k_1 \k_2 \k_3 \k_4}\right) &\propto
    \Bigg(\sum_{n=0}^{+\infty}\frac{(-1)^n u^{d-2}}{\left(n+\frac{d-2}{2}\right)^2+\mu_\chi^2}\left(\frac{u}{v}\right)^{n}\frac{\Gamma(n+d-2)}{\Gamma(n+1)}\,_2F_1\left(
    \begin{matrix}
        \frac{1-n}{2}\,,\,-\frac{n}{2} \\ -n-\frac{d-4}{2}
    \end{matrix}\,,v^2\right)
    \,_2F_1\left(
    \begin{matrix}
        \frac{n+d-1}{2}\,,\,\frac{n+d-2}{2} \\ n+\frac{d}{2}
    \end{matrix}\,,u^2\right)\\
    &- \frac{1}{2}\left(u^{-2}-1\right)^{\frac{3-d}{4}}
    \left(v^{-2}-1\right)^{\frac{3-d}{4}}\left(\Gamma\left(\frac{d-2}{2}\pm i\mu_\chi\right)\right)^2 P_{i\mu_\chi-\frac{1}{2}}^{-\frac{d-3}{2}}(e^{+i\pi}u^{-1})\,P_{i\mu_\chi-\frac{1}{2}}^{-\frac{d-3}{2}}(v^{-1})\Bigg)\,,
    \end{aligned}
\end{equation}    
\end{widetext}
where $u, v\equiv k_I/k_{12, 34}$ (here $u<v$), $k_{ij}\equiv |\k_i| + |\k_j|$, and $\,_2F_1$ and $P_{i\mu-1/2}^{-\frac{d-3}{2}}$ are, respectively, hypergeometric and associated Legendre functions.


\vskip 4pt
{\bf Spectral density and loop momentum integrals.} 
At the nonperturbative level, the Wightman two-point function of any scalar operator $\O$ can be expressed in the K\"all\'en-Lehmann spectral representation~\cite{Bros:1990cu,Bros:1995js,DiPietro:2021sjt,Hogervorst:2021uvp,Loparco:2023rug} by inserting the resolution of the identity in the momentum space basis $\ket{\mu,\k}$:
\begin{equation}
\label{eq: KLspectrum}
    \begin{aligned}
     &\bra{\Omega}\O_{-\aa}(z_1,\x_1)\O_{\aa}(z_2,\x_2)\ket{\Omega}\\&=\int_{\P\oplus\C}\d\mu\frac{\d^d\k}{(2\pi)^d}\rho^{\P,\C}_\O(\mu)\Phi^{\mu}_{\k}(z_1,\x_1)\Phi^{\mu}_{-\k}(z_2,\x_2) \,,
    \end{aligned}
\end{equation}
where the spectral density, which reads $\rho^{\P}_\O(\mu) = \N_{\mu} |c_\O(\mu)|^2$ for the principal series, is positive, as implied by unitarity.
For local and causal scalar theories in the Bunch-Davies vacuum, the exceptional series does not arise in the decomposition~\eqref{eq: KLspectrum}~\cite{Loparco:2023akg}. The complementary series contribution can be systematically obtained from the non-analyticities of $\rho^\P_\O(\mu)$, which is a meromorphic function in all known examples, see, e.g.~\cite{Loparco:2023rug}. Using the generalized delta function defined in~\eqref{eq:generalized-delta}, one can thus restrict the decomposition~\eqref{eq: KLspectrum} to KLF space, i.e.~real $\mu$, by writing $\rho_\O(\mu)=\rho^\P_\O(\mu) + 4 i \pi \sum_n {\textrm{Res} (\rho_{\O}^{\P}},\mu_n) \hat{\delta}_{\mu_n}(\mu)$, where $\mu_n$ are the poles of $\rho^\P_\O(\mu)$ in $(-i(d/2-\Delta),i(d/2-\Delta))$ with $\Delta$ being the leading behavior of the two-point function at large separation~\cite{Loparco:2023akg,unpublishedkey}.
The spectral density $\rho_\O(\mu)$ relates the nonperturbative two-point function to free propagators~\eqref{eq: propagator matrix} through 
\begin{equation}
    \bm{\G}_\O(\mu) = \int \displaylimits_{-\infty}^{+\infty} \d\mut \rho_\O(\mut) \bm{\G}_{\mut}(\mu) \,,
    \label{KLF-KL}
\end{equation}
where we defined the reduced two-point function matrix as $\boldsymbol{\mathcal{G}}_{\O}\left(^{\mu\,\mu'}_{\k\,\k'}\right) \equiv H^{-2} \delta(^{\mu\,\mu'}_{\k\,-\k'})\boldsymbol{\mathcal{G}}_{\O}(\mu)$. Eq.~\eqref{KLF-KL} represents the KLF form of the K\"all\'en-Lehmann representation, where off-diagonal terms reduce to $\rho_\O(\mu)=\N_{\mu} \G^{\O}_{-\aa\aa}(\mu)$. Consequently, the spectral density can be found by taking the inverse KLF transform of the Wightman function:
\begin{equation}
\label{eq: inversion formula}
    \begin{aligned}
        \frac{\rho_\O(\mu)}{\N_\mu} = \int \displaylimits_{\EAdS'} &\frac{\Phi_{\k}^{\mu}(z', \x')}{\Phi^\mu_{-\k}(z,\x)}\braket{\Omega|\O_{-\a}(z', \x')\O_{\a}(z, \x)|\Omega}\,.
    \end{aligned}
\end{equation}
Eventually, the group-theoretical structure of our construction enables us to streamline and give a transparent physical meaning to otherwise obscure identities. As a simple example, one-loop corrections to the scalar propagator include momentum integrals involving two cubic vertex functions $\I^{\mu_1 \mu_2 \mu}_{k_1 k_2 k}$. The latter are directly related to the decomposition of the tensor product of the UIRs $\mu_1$ and $\mu_2$ in the KLF space:
\begin{equation}
\label{eq: Expression CG coef}
   \begin{aligned}
    c_{\varphi_1\varphi_2}(\mu)  \left(\begin{array}{crc}
          \mu_1 & \mu_2 & \mu\\
           \k_1& \k_2 & \k
      \end{array}\right)^* &= (2\pi)^d\delta^{(d)}(\k_1+\k_2-\k)\\
       &\times c_{\varphi_1}(\mu_1)c_{\varphi_2}(\mu_2)\frac{H^{\frac{d+1}{2}}}{\pi^\frac{3}{2}}\I^{\mu_1\mu_2\mu}_{k_1k_2k} \,,
  \end{aligned}
\end{equation}
where we defined the dS 3-$\mu$ symbols as $ \left(\begin{array}{crc}
          \mu_1 & \mu_2 & \mu\\
           \k_1& \k_2 & \k
\end{array}\right)\equiv \left(\bra{\mu_1,\k_1}\otimes\bra{\mu_2,\k_2}\right) \ket{\mu,\k}$ and $\varphi_{1,2}$ are free principal series fields with frequencies $\mu_{1,2}$. As such, they obey the simple orthogonality relation:
\begin{equation}\label{eq: orthogonality relations C}
    \delta\left(^{\mu\, \mu'}_{\k\,\k'}\right)=\frac{\N_{\mu_1}\N_{\mu_2}}{\left[\delta(0)\right]^2}\int\displaylimits_{\k_1,\k_2}\left(\begin{array}{crc}
        \mu_1 & \mu_2 & \mu \\
        \k_1 & \k_2 & \k
    \end{array}\right)^*\left(\begin{array}{crc}
        \mu_1 & \mu_2 & \mu' \\
        \k_1 & \k_2 & \k'
    \end{array}\right)\;.
\end{equation}
This can be obtained by inserting a resolution of the identity within the orthogonality relation $\braket{\mu,\k|\mu',\k'}=\delta(^{\mu\,\mu'}_{\k\,\k'})$. Using the expression~\eqref{eq: Expression CG coef} and $\rho_{\O}(\mu)={\cal N}_\mu |c_\O(\mu)|^2$ yields the following expression for the loop integral:
\begin{equation}
    \int\frac{\d^d\q}{(2\pi)^d}\I^{\mu \mu_{\varphi_1}\mu_{\varphi_2}}_{k q|\k-\q|}\I^{\mu' \mu_{\varphi_1}\mu_{\varphi_2}}_{k q|\k-\q|}=\frac{\pi^3}{H^{d-1}}\frac{\hat{\delta}_{\mu'}(\mu)}{\N_{\mu}^2}\rho_{\varphi_1\varphi_2}(\mu)\;,
\end{equation}
where $\rho_{\varphi_1\varphi_2}(\mu)$, the spectral density of the composite operator $\O=\varphi_1\varphi_2$, has been computed in~\cite{Bros:2009bz,Marolf:2010zp} and more recently in~\cite{Sleight:2021plv,Loparco:2023rug}.


\vskip 4pt
{\bf Conclusion and outlook.} 
In this letter, we have introduced a novel, fully nonperturbative momentum-space formulation of QFT in the Poincar\'e patch of dS spacetime, dual to configuration space. This framework enables a reformulation of QFT directly in this space, conveniently defined on the EAdS SK double-branch path-integral contour. It also provides a straightforward correspondence between cosmological and amputated KLF correlators.

This new language makes several key features manifest: (i) the nonconservation of the dS frequency, (ii) the natural emergence of the Bessel-type dS harmonic function for the time/frequency dual correspondence, and (iii) a reorganization of perturbative calculations wherein time ordering is replaced by spectral integration. As a result, the underlying structure of cosmological correlators becomes transparent, and the latter are cast in a form that facilitates contact with flat space.

There are many questions directly spurred by our foundational construction. For example, it is well motivated to ask to what extent KLF space can be generalized or deformed to accommodate controlled (and perturbative) symmetry breaking of the dS isometries. Since a convenient byproduct of our construction is that frequency integration appears naturally and spectral density can be readily extracted, it would be very interesting to use this tool to dress propagators through resummation. Finally, we also leave for future work the possibility to connect to flat-space amplitudes, where energy conservation emerges from spectral integration. These and related lines of investigation hold the promise of fundamental new insights into understanding QFT in dS.


\vskip 4pt
{\bf Acknowledgments.} We thank
Cliff Burgess, Thomas Colas, Guillaume Faye, Francesco Nitti, Nadine Nussbaumer, Piotr Tourkine, and Pierre Vanhove for useful discussions. The research of DW is funded by the European Union (ERC, \raisebox{-2pt}{\includegraphics[height=0.9\baselineskip]{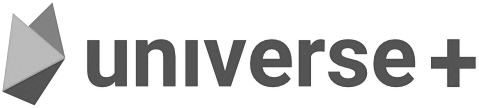}}, 101118787). Views and opinions expressed are however those of the author(s) only and do not necessarily reflect those of the European Union or the European Research Council Executive Agency. Neither the European Union nor the granting authority can be held responsible for them.

\bibliography{Bibliography}

\begin{thebibliography}{38}%
\makeatletter
\providecommand \@ifxundefined [1]{%
 \@ifx{#1\undefined}
}%
\providecommand \@ifnum [1]{%
 \ifnum #1\expandafter \@firstoftwo
 \else \expandafter \@secondoftwo
 \fi
}%
\providecommand \@ifx [1]{%
 \ifx #1\expandafter \@firstoftwo
 \else \expandafter \@secondoftwo
 \fi
}%
\providecommand \natexlab [1]{#1}%
\providecommand \enquote  [1]{``#1''}%
\providecommand \bibnamefont  [1]{#1}%
\providecommand \bibfnamefont [1]{#1}%
\providecommand \citenamefont [1]{#1}%
\providecommand \href@noop [0]{\@secondoftwo}%
\providecommand \href [0]{\begingroup \@sanitize@url \@href}%
\providecommand \@href[1]{\@@startlink{#1}\@@href}%
\providecommand \@@href[1]{\endgroup#1\@@endlink}%
\providecommand \@sanitize@url [0]{\catcode `\\12\catcode `\$12\catcode
  `\&12\catcode `\#12\catcode `\^12\catcode `\_12\catcode `\%12\relax}%
\providecommand \@@startlink[1]{}%
\providecommand \@@endlink[0]{}%
\providecommand \url  [0]{\begingroup\@sanitize@url \@url }%
\providecommand \@url [1]{\endgroup\@href {#1}{\urlprefix }}%
\providecommand \urlprefix  [0]{URL }%
\providecommand \Eprint [0]{\href }%
\providecommand \doibase [0]{http://dx.doi.org/}%
\providecommand \selectlanguage [0]{\@gobble}%
\providecommand \bibinfo  [0]{\@secondoftwo}%
\providecommand \bibfield  [0]{\@secondoftwo}%
\providecommand \translation [1]{[#1]}%
\providecommand \BibitemOpen [0]{}%
\providecommand \bibitemStop [0]{}%
\providecommand \bibitemNoStop [0]{.\EOS\space}%
\providecommand \EOS [0]{\spacefactor3000\relax}%
\providecommand \BibitemShut  [1]{\csname bibitem#1\endcsname}%
\let\auto@bib@innerbib\@empty
\bibitem [{\citenamefont {Strominger}(2001)}]{Strominger:2001pn}%
  \BibitemOpen
  \bibfield  {author} {\bibinfo {author} {\bibfnamefont {A.}~\bibnamefont
  {Strominger}},\ }\href {\doibase 10.1088/1126-6708/2001/10/034} {\bibfield
  {journal} {\bibinfo  {journal} {JHEP}\ }\textbf {\bibinfo {volume} {10}},\
  \bibinfo {pages} {034} (\bibinfo {year} {2001})},\ \Eprint
  {http://arxiv.org/abs/hep-th/0106113} {arXiv:hep-th/0106113} \BibitemShut
  {NoStop}%
\bibitem [{\citenamefont {Witten}(2001)}]{Witten:2001kn}%
  \BibitemOpen
  \bibfield  {author} {\bibinfo {author} {\bibfnamefont {E.}~\bibnamefont
  {Witten}},\ }\bibfield  {booktitle} {\emph {\bibinfo {booktitle} {{Strings
  2001: International Conference}}},\ }\href@noop {} {\  (\bibinfo {year}
  {2001})},\ \Eprint {http://arxiv.org/abs/hep-th/0106109}
  {arXiv:hep-th/0106109} \BibitemShut {NoStop}%
\bibitem [{\citenamefont {Marolf}\ \emph {et~al.}(2013)\citenamefont {Marolf},
  \citenamefont {Morrison},\ and\ \citenamefont {Srednicki}}]{Marolf:2012kh}%
  \BibitemOpen
  \bibfield  {author} {\bibinfo {author} {\bibfnamefont {D.}~\bibnamefont
  {Marolf}}, \bibinfo {author} {\bibfnamefont {I.~A.}\ \bibnamefont
  {Morrison}}, \ and\ \bibinfo {author} {\bibfnamefont {M.}~\bibnamefont
  {Srednicki}},\ }\href {\doibase 10.1088/0264-9381/30/15/155023} {\bibfield
  {journal} {\bibinfo  {journal} {Class. Quant. Grav.}\ }\textbf {\bibinfo
  {volume} {30}},\ \bibinfo {pages} {155023} (\bibinfo {year} {2013})},\
  \Eprint {http://arxiv.org/abs/1209.6039} {arXiv:1209.6039 [hep-th]}
  \BibitemShut {NoStop}%
\bibitem [{\citenamefont {Melville}\ and\ \citenamefont
  {Pimentel}(2024{\natexlab{a}})}]{Melville:2023kgd}%
  \BibitemOpen
  \bibfield  {author} {\bibinfo {author} {\bibfnamefont {S.}~\bibnamefont
  {Melville}}\ and\ \bibinfo {author} {\bibfnamefont {G.~L.}\ \bibnamefont
  {Pimentel}},\ }\href {\doibase 10.1103/PhysRevD.110.103530} {\bibfield
  {journal} {\bibinfo  {journal} {Phys. Rev. D}\ }\textbf {\bibinfo {volume}
  {110}},\ \bibinfo {pages} {103530} (\bibinfo {year} {2024}{\natexlab{a}})},\
  \Eprint {http://arxiv.org/abs/2309.07092} {arXiv:2309.07092 [hep-th]}
  \BibitemShut {NoStop}%
\bibitem [{\citenamefont {Donath}\ and\ \citenamefont
  {Pajer}(2024)}]{Donath:2024utn}%
  \BibitemOpen
  \bibfield  {author} {\bibinfo {author} {\bibfnamefont {Y.}~\bibnamefont
  {Donath}}\ and\ \bibinfo {author} {\bibfnamefont {E.}~\bibnamefont {Pajer}},\
  }\href {\doibase 10.1007/JHEP07(2024)064} {\bibfield  {journal} {\bibinfo
  {journal} {JHEP}\ }\textbf {\bibinfo {volume} {07}},\ \bibinfo {pages} {064}
  (\bibinfo {year} {2024})},\ \Eprint {http://arxiv.org/abs/2402.05999}
  {arXiv:2402.05999 [hep-th]} \BibitemShut {NoStop}%
\bibitem [{\citenamefont {Melville}\ and\ \citenamefont
  {Pimentel}(2024{\natexlab{b}})}]{Melville:2024ove}%
  \BibitemOpen
  \bibfield  {author} {\bibinfo {author} {\bibfnamefont {S.}~\bibnamefont
  {Melville}}\ and\ \bibinfo {author} {\bibfnamefont {G.~L.}\ \bibnamefont
  {Pimentel}},\ }\href {\doibase 10.1007/JHEP08(2024)211} {\bibfield  {journal}
  {\bibinfo  {journal} {JHEP}\ }\textbf {\bibinfo {volume} {08}},\ \bibinfo
  {pages} {211} (\bibinfo {year} {2024}{\natexlab{b}})},\ \Eprint
  {http://arxiv.org/abs/2404.05712} {arXiv:2404.05712 [hep-th]} \BibitemShut
  {NoStop}%
\bibitem [{\citenamefont {Di~Pietro}\ \emph {et~al.}(2022)\citenamefont
  {Di~Pietro}, \citenamefont {Gorbenko},\ and\ \citenamefont
  {Komatsu}}]{DiPietro:2021sjt}%
  \BibitemOpen
  \bibfield  {author} {\bibinfo {author} {\bibfnamefont {L.}~\bibnamefont
  {Di~Pietro}}, \bibinfo {author} {\bibfnamefont {V.}~\bibnamefont {Gorbenko}},
  \ and\ \bibinfo {author} {\bibfnamefont {S.}~\bibnamefont {Komatsu}},\ }\href
  {\doibase 10.1007/JHEP03(2022)023} {\bibfield  {journal} {\bibinfo  {journal}
  {JHEP}\ }\textbf {\bibinfo {volume} {03}},\ \bibinfo {pages} {023} (\bibinfo
  {year} {2022})},\ \Eprint {http://arxiv.org/abs/2108.01695} {arXiv:2108.01695
  [hep-th]} \BibitemShut {NoStop}%
\bibitem [{\citenamefont {Xianyu}\ and\ \citenamefont
  {Zhang}(2023)}]{Xianyu:2022jwk}%
  \BibitemOpen
  \bibfield  {author} {\bibinfo {author} {\bibfnamefont {Z.-Z.}\ \bibnamefont
  {Xianyu}}\ and\ \bibinfo {author} {\bibfnamefont {H.}~\bibnamefont {Zhang}},\
  }\href {\doibase 10.1007/JHEP04(2023)103} {\bibfield  {journal} {\bibinfo
  {journal} {JHEP}\ }\textbf {\bibinfo {volume} {04}},\ \bibinfo {pages} {103}
  (\bibinfo {year} {2023})},\ \Eprint {http://arxiv.org/abs/2211.03810}
  {arXiv:2211.03810 [hep-th]} \BibitemShut {NoStop}%
\bibitem [{\citenamefont {Chakraborty}\ and\ \citenamefont
  {Stout}(2024)}]{Chakraborty:2023qbp}%
  \BibitemOpen
  \bibfield  {author} {\bibinfo {author} {\bibfnamefont {P.}~\bibnamefont
  {Chakraborty}}\ and\ \bibinfo {author} {\bibfnamefont {J.}~\bibnamefont
  {Stout}},\ }\href {\doibase 10.1007/JHEP02(2024)021} {\bibfield  {journal}
  {\bibinfo  {journal} {JHEP}\ }\textbf {\bibinfo {volume} {02}},\ \bibinfo
  {pages} {021} (\bibinfo {year} {2024})},\ \Eprint
  {http://arxiv.org/abs/2310.01494} {arXiv:2310.01494 [hep-th]} \BibitemShut
  {NoStop}%
\bibitem [{\citenamefont {Qin}\ and\ \citenamefont
  {Xianyu}(2023)}]{Qin:2023bjk}%
  \BibitemOpen
  \bibfield  {author} {\bibinfo {author} {\bibfnamefont {Z.}~\bibnamefont
  {Qin}}\ and\ \bibinfo {author} {\bibfnamefont {Z.-Z.}\ \bibnamefont
  {Xianyu}},\ }\href {\doibase 10.1007/JHEP09(2023)116} {\bibfield  {journal}
  {\bibinfo  {journal} {JHEP}\ }\textbf {\bibinfo {volume} {09}},\ \bibinfo
  {pages} {116} (\bibinfo {year} {2023})},\ \Eprint
  {http://arxiv.org/abs/2304.13295} {arXiv:2304.13295 [hep-th]} \BibitemShut
  {NoStop}%
\bibitem [{\citenamefont {Qin}\ and\ \citenamefont
  {Xianyu}(2024)}]{Qin:2023nhv}%
  \BibitemOpen
  \bibfield  {author} {\bibinfo {author} {\bibfnamefont {Z.}~\bibnamefont
  {Qin}}\ and\ \bibinfo {author} {\bibfnamefont {Z.-Z.}\ \bibnamefont
  {Xianyu}},\ }\href {\doibase 10.1007/JHEP01(2024)168} {\bibfield  {journal}
  {\bibinfo  {journal} {JHEP}\ }\textbf {\bibinfo {volume} {01}},\ \bibinfo
  {pages} {168} (\bibinfo {year} {2024})},\ \Eprint
  {http://arxiv.org/abs/2308.14802} {arXiv:2308.14802 [hep-th]} \BibitemShut
  {NoStop}%
\bibitem [{\citenamefont {Chen}\ \emph {et~al.}(2017)\citenamefont {Chen},
  \citenamefont {Wang},\ and\ \citenamefont {Xianyu}}]{Chen:2017ryl}%
  \BibitemOpen
  \bibfield  {author} {\bibinfo {author} {\bibfnamefont {X.}~\bibnamefont
  {Chen}}, \bibinfo {author} {\bibfnamefont {Y.}~\bibnamefont {Wang}}, \ and\
  \bibinfo {author} {\bibfnamefont {Z.-Z.}\ \bibnamefont {Xianyu}},\ }\href
  {\doibase 10.1088/1475-7516/2017/12/006} {\bibfield  {journal} {\bibinfo
  {journal} {JCAP}\ }\textbf {\bibinfo {volume} {12}},\ \bibinfo {pages} {006}
  (\bibinfo {year} {2017})},\ \Eprint {http://arxiv.org/abs/1703.10166}
  {arXiv:1703.10166 [hep-th]} \BibitemShut {NoStop}%
\bibitem [{\citenamefont {Higuchi}(1987)}]{Higuchi:1986wu}%
  \BibitemOpen
  \bibfield  {author} {\bibinfo {author} {\bibfnamefont {A.}~\bibnamefont
  {Higuchi}},\ }\href {\doibase 10.1063/1.527513} {\bibfield  {journal}
  {\bibinfo  {journal} {J. Math. Phys.}\ }\textbf {\bibinfo {volume} {28}},\
  \bibinfo {pages} {1553} (\bibinfo {year} {1987})},\ \bibinfo {note}
  {[Erratum: J.Math.Phys. 43, 6385 (2002)]}\BibitemShut {NoStop}%
\bibitem [{\citenamefont {Bros}(1991)}]{Bros:1990cu}%
  \BibitemOpen
  \bibfield  {author} {\bibinfo {author} {\bibfnamefont {J.}~\bibnamefont
  {Bros}},\ }\href {\doibase 10.1016/0920-5632(91)90119-Y} {\bibfield
  {journal} {\bibinfo  {journal} {Nucl. Phys. B Proc. Suppl.}\ }\textbf
  {\bibinfo {volume} {18}},\ \bibinfo {pages} {22} (\bibinfo {year}
  {1991})}\BibitemShut {NoStop}%
\bibitem [{\citenamefont {Bros}\ and\ \citenamefont
  {Moschella}(1996)}]{Bros:1995js}%
  \BibitemOpen
  \bibfield  {author} {\bibinfo {author} {\bibfnamefont {J.}~\bibnamefont
  {Bros}}\ and\ \bibinfo {author} {\bibfnamefont {U.}~\bibnamefont
  {Moschella}},\ }\href {\doibase 10.1142/S0129055X96000123} {\bibfield
  {journal} {\bibinfo  {journal} {Rev. Math. Phys.}\ }\textbf {\bibinfo
  {volume} {8}},\ \bibinfo {pages} {327} (\bibinfo {year} {1996})},\ \Eprint
  {http://arxiv.org/abs/gr-qc/9511019} {arXiv:gr-qc/9511019} \BibitemShut
  {NoStop}%
\bibitem [{\citenamefont {Marolf}\ and\ \citenamefont
  {Morrison}(2010)}]{Marolf:2010zp}%
  \BibitemOpen
  \bibfield  {author} {\bibinfo {author} {\bibfnamefont {D.}~\bibnamefont
  {Marolf}}\ and\ \bibinfo {author} {\bibfnamefont {I.~A.}\ \bibnamefont
  {Morrison}},\ }\href {\doibase 10.1103/PhysRevD.82.105032} {\bibfield
  {journal} {\bibinfo  {journal} {Phys. Rev. D}\ }\textbf {\bibinfo {volume}
  {82}},\ \bibinfo {pages} {105032} (\bibinfo {year} {2010})},\ \Eprint
  {http://arxiv.org/abs/1006.0035} {arXiv:1006.0035 [gr-qc]} \BibitemShut
  {NoStop}%
\bibitem [{\citenamefont {Marolf}\ and\ \citenamefont
  {Morrison}(2011)}]{Marolf:2010nz}%
  \BibitemOpen
  \bibfield  {author} {\bibinfo {author} {\bibfnamefont {D.}~\bibnamefont
  {Marolf}}\ and\ \bibinfo {author} {\bibfnamefont {I.~A.}\ \bibnamefont
  {Morrison}},\ }\href {\doibase 10.1103/PhysRevD.84.044040} {\bibfield
  {journal} {\bibinfo  {journal} {Phys. Rev. D}\ }\textbf {\bibinfo {volume}
  {84}},\ \bibinfo {pages} {044040} (\bibinfo {year} {2011})},\ \Eprint
  {http://arxiv.org/abs/1010.5327} {arXiv:1010.5327 [gr-qc]} \BibitemShut
  {NoStop}%
\bibitem [{\citenamefont {Higuchi}\ \emph {et~al.}(2011)\citenamefont
  {Higuchi}, \citenamefont {Marolf},\ and\ \citenamefont
  {Morrison}}]{Higuchi:2010xt}%
  \BibitemOpen
  \bibfield  {author} {\bibinfo {author} {\bibfnamefont {A.}~\bibnamefont
  {Higuchi}}, \bibinfo {author} {\bibfnamefont {D.}~\bibnamefont {Marolf}}, \
  and\ \bibinfo {author} {\bibfnamefont {I.~A.}\ \bibnamefont {Morrison}},\
  }\href {\doibase 10.1103/PhysRevD.83.084029} {\bibfield  {journal} {\bibinfo
  {journal} {Phys. Rev. D}\ }\textbf {\bibinfo {volume} {83}},\ \bibinfo
  {pages} {084029} (\bibinfo {year} {2011})},\ \Eprint
  {http://arxiv.org/abs/1012.3415} {arXiv:1012.3415 [gr-qc]} \BibitemShut
  {NoStop}%
\bibitem [{\citenamefont {Belrhali}\ \emph {et~al.}(2026)\citenamefont
  {Belrhali}, \citenamefont {Poisson}, \citenamefont {Renaux-Petel},\ and\
  \citenamefont {Werth}}]{Belrhali:2026rkn}%
  \BibitemOpen
  \bibfield  {author} {\bibinfo {author} {\bibfnamefont {N.}~\bibnamefont
  {Belrhali}}, \bibinfo {author} {\bibfnamefont {A.}~\bibnamefont {Poisson}},
  \bibinfo {author} {\bibfnamefont {S.}~\bibnamefont {Renaux-Petel}}, \ and\
  \bibinfo {author} {\bibfnamefont {D.}~\bibnamefont {Werth}},\ }\href@noop {}
  {\  (\bibinfo {year} {2026})},\ \Eprint {http://arxiv.org/abs/2604.15251}
  {arXiv:2604.15251 [hep-th]} \BibitemShut {NoStop}%
\bibitem [{\citenamefont {Bargmann}\ and\ \citenamefont
  {Wigner}(1948)}]{Bargmann:1948ck}%
  \BibitemOpen
  \bibfield  {author} {\bibinfo {author} {\bibfnamefont {V.}~\bibnamefont
  {Bargmann}}\ and\ \bibinfo {author} {\bibfnamefont {E.~P.}\ \bibnamefont
  {Wigner}},\ }\href {\doibase 10.1073/pnas.34.5.211} {\bibfield  {journal}
  {\bibinfo  {journal} {Proc. Nat. Acad. Sci.}\ }\textbf {\bibinfo {volume}
  {34}},\ \bibinfo {pages} {211} (\bibinfo {year} {1948})}\BibitemShut
  {NoStop}%
\bibitem [{\citenamefont {Sun}(2025)}]{Sun:2021thf}%
  \BibitemOpen
  \bibfield  {author} {\bibinfo {author} {\bibfnamefont {Z.}~\bibnamefont
  {Sun}},\ }\href {\doibase 10.1142/S0129055X24300073} {\bibfield  {journal}
  {\bibinfo  {journal} {Rev. Math. Phys.}\ }\textbf {\bibinfo {volume} {37}},\
  \bibinfo {pages} {2430007} (\bibinfo {year} {2025})},\ \Eprint
  {http://arxiv.org/abs/2111.04591} {arXiv:2111.04591 [hep-th]} \BibitemShut
  {NoStop}%
\bibitem [{\citenamefont {Bonifacio}\ \emph {et~al.}(2019)\citenamefont
  {Bonifacio}, \citenamefont {Hinterbichler}, \citenamefont {Joyce},\ and\
  \citenamefont {Rosen}}]{Bonifacio:2018zex}%
  \BibitemOpen
  \bibfield  {author} {\bibinfo {author} {\bibfnamefont {J.}~\bibnamefont
  {Bonifacio}}, \bibinfo {author} {\bibfnamefont {K.}~\bibnamefont
  {Hinterbichler}}, \bibinfo {author} {\bibfnamefont {A.}~\bibnamefont
  {Joyce}}, \ and\ \bibinfo {author} {\bibfnamefont {R.~A.}\ \bibnamefont
  {Rosen}},\ }\href {\doibase 10.1007/JHEP02(2019)178} {\bibfield  {journal}
  {\bibinfo  {journal} {JHEP}\ }\textbf {\bibinfo {volume} {02}},\ \bibinfo
  {pages} {178} (\bibinfo {year} {2019})},\ \Eprint
  {http://arxiv.org/abs/1812.08167} {arXiv:1812.08167 [hep-th]} \BibitemShut
  {NoStop}%
\bibitem [{\citenamefont {Weinberg}(2005)}]{Weinberg:2005vy}%
  \BibitemOpen
  \bibfield  {author} {\bibinfo {author} {\bibfnamefont {S.}~\bibnamefont
  {Weinberg}},\ }\href {\doibase 10.1103/PhysRevD.72.043514} {\bibfield
  {journal} {\bibinfo  {journal} {Phys. Rev. D}\ }\textbf {\bibinfo {volume}
  {72}},\ \bibinfo {pages} {043514} (\bibinfo {year} {2005})},\ \Eprint
  {http://arxiv.org/abs/hep-th/0506236} {arXiv:hep-th/0506236} \BibitemShut
  {NoStop}%
\bibitem [{\citenamefont {Anninos}\ \emph {et~al.}(2015)\citenamefont
  {Anninos}, \citenamefont {Anous}, \citenamefont {Freedman},\ and\
  \citenamefont {Konstantinidis}}]{Anninos:2014lwa}%
  \BibitemOpen
  \bibfield  {author} {\bibinfo {author} {\bibfnamefont {D.}~\bibnamefont
  {Anninos}}, \bibinfo {author} {\bibfnamefont {T.}~\bibnamefont {Anous}},
  \bibinfo {author} {\bibfnamefont {D.~Z.}\ \bibnamefont {Freedman}}, \ and\
  \bibinfo {author} {\bibfnamefont {G.}~\bibnamefont {Konstantinidis}},\ }\href
  {\doibase 10.1088/1475-7516/2015/11/048} {\bibfield  {journal} {\bibinfo
  {journal} {JCAP}\ }\textbf {\bibinfo {volume} {11}},\ \bibinfo {pages} {048}
  (\bibinfo {year} {2015})},\ \Eprint {http://arxiv.org/abs/1406.5490}
  {arXiv:1406.5490 [hep-th]} \BibitemShut {NoStop}%
\bibitem [{\citenamefont {Sleight}\ and\ \citenamefont
  {Taronna}(2021)}]{Sleight:2021plv}%
  \BibitemOpen
  \bibfield  {author} {\bibinfo {author} {\bibfnamefont {C.}~\bibnamefont
  {Sleight}}\ and\ \bibinfo {author} {\bibfnamefont {M.}~\bibnamefont
  {Taronna}},\ }\href {\doibase 10.1007/JHEP12(2021)074} {\bibfield  {journal}
  {\bibinfo  {journal} {JHEP}\ }\textbf {\bibinfo {volume} {12}},\ \bibinfo
  {pages} {074} (\bibinfo {year} {2021})},\ \Eprint
  {http://arxiv.org/abs/2109.02725} {arXiv:2109.02725 [hep-th]} \BibitemShut
  {NoStop}%
\bibitem [{\citenamefont {Loparco}\ \emph {et~al.}(2023)\citenamefont
  {Loparco}, \citenamefont {Penedones}, \citenamefont {Salehi~Vaziri},\ and\
  \citenamefont {Sun}}]{Loparco:2023rug}%
  \BibitemOpen
  \bibfield  {author} {\bibinfo {author} {\bibfnamefont {M.}~\bibnamefont
  {Loparco}}, \bibinfo {author} {\bibfnamefont {J.}~\bibnamefont {Penedones}},
  \bibinfo {author} {\bibfnamefont {K.}~\bibnamefont {Salehi~Vaziri}}, \ and\
  \bibinfo {author} {\bibfnamefont {Z.}~\bibnamefont {Sun}},\ }\href {\doibase
  10.1007/JHEP12(2023)159} {\bibfield  {journal} {\bibinfo  {journal} {JHEP}\
  }\textbf {\bibinfo {volume} {12}},\ \bibinfo {pages} {159} (\bibinfo {year}
  {2023})},\ \Eprint {http://arxiv.org/abs/2306.00090} {arXiv:2306.00090
  [hep-th]} \BibitemShut {NoStop}%
\bibitem [{\citenamefont {Chowdhury}\ \emph {et~al.}(2025)\citenamefont
  {Chowdhury}, \citenamefont {Lipstein}, \citenamefont {Mei}, \citenamefont
  {Sachs},\ and\ \citenamefont {Vanhove}}]{Chowdhury:2023arc}%
  \BibitemOpen
  \bibfield  {author} {\bibinfo {author} {\bibfnamefont {C.}~\bibnamefont
  {Chowdhury}}, \bibinfo {author} {\bibfnamefont {A.}~\bibnamefont {Lipstein}},
  \bibinfo {author} {\bibfnamefont {J.}~\bibnamefont {Mei}}, \bibinfo {author}
  {\bibfnamefont {I.}~\bibnamefont {Sachs}}, \ and\ \bibinfo {author}
  {\bibfnamefont {P.}~\bibnamefont {Vanhove}},\ }\href {\doibase
  10.1007/JHEP03(2025)007} {\bibfield  {journal} {\bibinfo  {journal} {JHEP}\
  }\textbf {\bibinfo {volume} {03}},\ \bibinfo {pages} {007} (\bibinfo {year}
  {2025})},\ \Eprint {http://arxiv.org/abs/2312.13803} {arXiv:2312.13803
  [hep-th]} \BibitemShut {NoStop}%
\bibitem [{{\relax DLMF}()}]{NIST:DLMF}%
  \BibitemOpen
  {\relax DLMF},\ \href {https://dlmf.nist.gov/} {\enquote {\bibinfo {title}
  {{\it NIST Digital Library of Mathematical Functions}},}\ }\bibinfo
  {howpublished} {\url{https://dlmf.nist.gov/}, Release 1.2.5 of 2025-12-15},\
  \bibinfo {note} {f.~W.~J. Olver, A.~B. {Olde Daalhuis}, D.~W. Lozier, B.~I.
  Schneider, R.~F. Boisvert, C.~W. Clark, B.~R. Miller, B.~V. Saunders, H.~S.
  Cohl, and M.~A. McClain, eds.}\BibitemShut {Stop}%
\bibitem [{\citenamefont {Kontorovich}\ and\ \citenamefont
  {Lebedev}(1938)}]{Kontorovich1938}%
  \BibitemOpen
  \bibfield  {author} {\bibinfo {author} {\bibfnamefont {M.}~\bibnamefont
  {Kontorovich}}\ and\ \bibinfo {author} {\bibfnamefont {N.}~\bibnamefont
  {Lebedev}},\ }\href@noop {} {\bibfield  {journal} {\bibinfo  {journal}
  {Doklady Akademii Nauk SSSR}\ }\textbf {\bibinfo {volume} {19}},\ \bibinfo
  {pages} {441} (\bibinfo {year} {1938})}\BibitemShut {NoStop}%
\bibitem [{\citenamefont {Lebedev}(1946)}]{Lebedev1946}%
  \BibitemOpen
  \bibfield  {author} {\bibinfo {author} {\bibfnamefont {N.}~\bibnamefont
  {Lebedev}},\ }\href@noop {} {\bibfield  {journal} {\bibinfo  {journal}
  {Doklady Akademii Nauk SSSR}\ }\textbf {\bibinfo {volume} {52}},\ \bibinfo
  {pages} {655} (\bibinfo {year} {1946})}\BibitemShut {NoStop}%
\bibitem [{\citenamefont {Gel'fand}\ and\ \citenamefont
  {Vilenkin}(1964)}]{Gelfand-Vilenkin}%
  \BibitemOpen
  \bibfield  {author} {\bibinfo {author} {\bibfnamefont {I.}~\bibnamefont
  {Gel'fand}}\ and\ \bibinfo {author} {\bibfnamefont {N.}~\bibnamefont
  {Vilenkin}},\ }\href {\doibase https://doi.org/10.1016%2Fc2013-0-12221-0}
  {\emph {\bibinfo {title} {{Generalized Functions, Vol. IV}}}}\ (\bibinfo
  {publisher} {Academic Press, New York},\ \bibinfo {year} {1964})\BibitemShut
  {NoStop}%
\bibitem [{\citenamefont {Maurin}(1968)}]{Maurin}%
  \BibitemOpen
  \bibfield  {author} {\bibinfo {author} {\bibfnamefont {K.}~\bibnamefont
  {Maurin}},\ }\href {https://api.semanticscholar.org/CorpusID:118845466} {\
  (\bibinfo {year} {1968})}\BibitemShut {NoStop}%
\bibitem [{\citenamefont {Werth}(2024)}]{Werth:2024mjg}%
  \BibitemOpen
  \bibfield  {author} {\bibinfo {author} {\bibfnamefont {D.}~\bibnamefont
  {Werth}},\ }\href {\doibase 10.1007/JHEP12(2024)017} {\bibfield  {journal}
  {\bibinfo  {journal} {JHEP}\ }\textbf {\bibinfo {volume} {12}},\ \bibinfo
  {pages} {017} (\bibinfo {year} {2024})},\ \Eprint
  {http://arxiv.org/abs/2409.02072} {arXiv:2409.02072 [hep-th]} \BibitemShut
  {NoStop}%
\bibitem [{\citenamefont {Bzowski}\ \emph {et~al.}(2014)\citenamefont
  {Bzowski}, \citenamefont {McFadden},\ and\ \citenamefont
  {Skenderis}}]{Bzowski:2013sza}%
  \BibitemOpen
  \bibfield  {author} {\bibinfo {author} {\bibfnamefont {A.}~\bibnamefont
  {Bzowski}}, \bibinfo {author} {\bibfnamefont {P.}~\bibnamefont {McFadden}}, \
  and\ \bibinfo {author} {\bibfnamefont {K.}~\bibnamefont {Skenderis}},\ }\href
  {\doibase 10.1007/JHEP03(2014)111} {\bibfield  {journal} {\bibinfo  {journal}
  {JHEP}\ }\textbf {\bibinfo {volume} {03}},\ \bibinfo {pages} {111} (\bibinfo
  {year} {2014})},\ \Eprint {http://arxiv.org/abs/1304.7760} {arXiv:1304.7760
  [hep-th]} \BibitemShut {NoStop}%
\bibitem [{\citenamefont {Bzowski}\ \emph {et~al.}(2016)\citenamefont
  {Bzowski}, \citenamefont {McFadden},\ and\ \citenamefont
  {Skenderis}}]{Bzowski:2015yxv}%
  \BibitemOpen
  \bibfield  {author} {\bibinfo {author} {\bibfnamefont {A.}~\bibnamefont
  {Bzowski}}, \bibinfo {author} {\bibfnamefont {P.}~\bibnamefont {McFadden}}, \
  and\ \bibinfo {author} {\bibfnamefont {K.}~\bibnamefont {Skenderis}},\ }\href
  {\doibase 10.1007/JHEP02(2016)068} {\bibfield  {journal} {\bibinfo  {journal}
  {JHEP}\ }\textbf {\bibinfo {volume} {02}},\ \bibinfo {pages} {068} (\bibinfo
  {year} {2016})},\ \Eprint {http://arxiv.org/abs/1511.02357} {arXiv:1511.02357
  [hep-th]} \BibitemShut {NoStop}%
\bibitem [{\citenamefont {Hogervorst}\ \emph {et~al.}(2023)\citenamefont
  {Hogervorst}, \citenamefont {Penedones},\ and\ \citenamefont
  {Vaziri}}]{Hogervorst:2021uvp}%
  \BibitemOpen
  \bibfield  {author} {\bibinfo {author} {\bibfnamefont {M.}~\bibnamefont
  {Hogervorst}}, \bibinfo {author} {\bibfnamefont {J.}~\bibnamefont
  {Penedones}}, \ and\ \bibinfo {author} {\bibfnamefont {K.~S.}\ \bibnamefont
  {Vaziri}},\ }\href {\doibase 10.1007/JHEP02(2023)162} {\bibfield  {journal}
  {\bibinfo  {journal} {JHEP}\ }\textbf {\bibinfo {volume} {02}},\ \bibinfo
  {pages} {162} (\bibinfo {year} {2023})},\ \Eprint
  {http://arxiv.org/abs/2107.13871} {arXiv:2107.13871 [hep-th]} \BibitemShut
  {NoStop}%
\bibitem [{\citenamefont {Loparco}\ \emph {et~al.}(2025)\citenamefont
  {Loparco}, \citenamefont {Qiao},\ and\ \citenamefont
  {Sun}}]{Loparco:2023akg}%
  \BibitemOpen
  \bibfield  {author} {\bibinfo {author} {\bibfnamefont {M.}~\bibnamefont
  {Loparco}}, \bibinfo {author} {\bibfnamefont {J.}~\bibnamefont {Qiao}}, \
  and\ \bibinfo {author} {\bibfnamefont {Z.}~\bibnamefont {Sun}},\ }\href
  {\doibase 10.21468/SciPostPhys.18.5.164} {\bibfield  {journal} {\bibinfo
  {journal} {SciPost Phys.}\ }\textbf {\bibinfo {volume} {18}},\ \bibinfo
  {pages} {164} (\bibinfo {year} {2025})},\ \Eprint
  {http://arxiv.org/abs/2310.15944} {arXiv:2310.15944 [hep-th]} \BibitemShut
  {NoStop}%
\bibitem [{\citenamefont {Bros}\ \emph {et~al.}(2010)\citenamefont {Bros},
  \citenamefont {Epstein}, \citenamefont {Gaudin}, \citenamefont {Moschella},\
  and\ \citenamefont {Pasquier}}]{Bros:2009bz}%
  \BibitemOpen
  \bibfield  {author} {\bibinfo {author} {\bibfnamefont {J.}~\bibnamefont
  {Bros}}, \bibinfo {author} {\bibfnamefont {H.}~\bibnamefont {Epstein}},
  \bibinfo {author} {\bibfnamefont {M.}~\bibnamefont {Gaudin}}, \bibinfo
  {author} {\bibfnamefont {U.}~\bibnamefont {Moschella}}, \ and\ \bibinfo
  {author} {\bibfnamefont {V.}~\bibnamefont {Pasquier}},\ }\href {\doibase
  10.1007/s00220-009-0875-4} {\bibfield  {journal} {\bibinfo  {journal}
  {Commun. Math. Phys.}\ }\textbf {\bibinfo {volume} {295}},\ \bibinfo {pages}
  {261} (\bibinfo {year} {2010})},\ \Eprint {http://arxiv.org/abs/0901.4223}
  {arXiv:0901.4223 [hep-th]} \BibitemShut {NoStop}%
\end{thebibliography}%
\end{document}